\documentclass[aps,prd,twocolumn,groupedaddress]{revtex4}

\usepackage{graphicx}% Include figure files

\begin{document}

%\preprint{}

\title{New results on catalyzed BBN with a long-lived negatively-charged
massive particle}


\author{Motohiko Kusakabe$^{1}$}
\email[]{kusakabe@icrr.u-tokyo.ac.jp}
\author{Toshitaka Kajino$^{2,3,4}$, Takashi Yoshida$^{3}$ and Grant J. Mathews$^{5}$}
%\email[]{kusakabe@icrr.u-tokyo.ac.jp}
%$\thanks{}
%\thanks{}

\affiliation{
$^1$Institute for Cosmic Ray Research, University of Tokyo, Kashiwa, Chiba 277-8582, Japan \\
$^2$Division of Theoretical Astronomy, National Astronomical Observatory
of Japan, Mitaka, Tokyo 181-8588, Japan \\
$^3$Department of Astronomy, Graduate School of Science, University of
Tokyo,  Hongo, Bunkyo-ku, Tokyo 113-0033, Japan \\
$^4$Department of Astronomical Science, The Graduate University for
Advanced Studies, Mitaka, Tokyo 181-8588, Japan \\
$^5$Department of Physics, Center for Astrophysics, University of
Notre Dame, Notre Dame, IN 46556, USA}


\date{\today}

\begin{abstract}

It has been proposed that the apparent discrepancies between the inferred  primordial abundances of $^6$Li and $^7$Li and the predictions of big bang
 nucleosynthesis (BBN) can be resolved by the existence of  a
 negatively-charged massive unstable supersymmetric particle ($X^-$) during the BBN epoch.   Here, we present new BBN  calculations with an $X^-$  particle utilizing  an improved nuclear reaction network
 including captures of nuclei by the particle, nuclear reactions and
 $\beta$-decays of normal nuclei and nuclei bound to the $X^-$ particles
 ($X$-nuclei), and new reaction rates derived from recent rigorous quantum many-body dynamical
 calculations.  We find  that this is still a viable model to explain the observed $^6$Li and $^7$Li abundances.  However, contrary to previous results,  neutral $X$-nuclei cannot significantly affect 
 the BBN light-element abundances.  We also show that with the new rates the production of heavier nuclei is suppressed and there is  no signature on abundances of nuclei heavier than Be in the
 $X^-$-particle  catalyzed BBN model as has been previously proposed.  We also consider   the version of this model whereby  the  $X^-$ particle decays into the present cold dark matter.  We analyze the this paradigm in light of the recent constraints on the dark-matter mass deduced from the possible detected events in the CDMS-II experiment.  We  conclude that  based upon the inferred range for the dark-matter mass, only  $X^-$  decay via the weak
 interaction can achieve  the desired $^7$Li destruction  while also reproducing the observed $^6$Li abundance.  

\end{abstract}

%26.35.+c Big Bang nucleosynthesis
%95.35.+d Dark matter
%98.80.Cq Particle-theory and field-theory models of the early Universe
%98.80.Es Observational cosmology (including Hubble constant, 
%          distance scale, cosmological constant, early Universe, etc)

%\keywords{}

\maketitle

%\section{Introduction}

%\section{Introduction}
The nucleosynthesis of light elements in the big bang is a unique  probe of new physics which may have occurred during the first few minutes of cosmic expansion in the big bang.  Of particular interest in this work is the apparent discrepancy between the inferred primordial  abundances of $^6$Li and $^7$Li and  the predictions of standard BBN.  A popular model to resolve this discrepancy is the existence of an unstable negatively charged supersymmetric particle during the nucleosynthesis epoch~\cite{Pospelov:2006sc,Kohri:2006cn,Cyburt:2006uv,Hamaguchi:2007mp,Bird:2007ge,Kusakabe:2007fu,Kusakabe:2007fv,Jedamzik_nega,Kamimura:2008fx,Pospelov:2007js,Kawasaki:2007xb,Jittoh_nega,Pospelov:2008ta}.  Depending upon their abundance and lifetime \cite{Kusakabe:2007fv}, such particles can catalyze the nuclear reactions leading to enhanced $^6$Li~\cite{Pospelov:2006sc} and depleted $^7$Li~\cite{Bird:2007ge,Kusakabe:2007fu} as required by observations.  Here we present new calculations based upon a substantially improved nuclear reaction network for this $X^-$-catalyzed BBN.   We solve numerically the
nonequilibrium nuclear and chemical reaction network associated to the $X^-$ particle~\cite{Kusakabe:2007fv}
with improved reaction rates derived from recent rigorous quantum many-body dynamical calculations~\cite{Kamimura:2008fx}.  We show that both the $^6$Li and
$^7$Li problems can still be solved.  However, contrary to earlier speculation \cite{Pospelov:2007js},  there is no signature in the
primordial abundances of heavier nuclides produced by this mechanism. 

Also in this work we examine  the version of this model in which the $X^-$ particles  decay into the present dark matter. In such models the allowed lifetimes and abundances can be  sensitive to the mass of the dark-matter particle.  In this regard the recent results  of the Cryogenic Dark Matter Search experiment (CDMS II) are of interest.  Possible detected events imply
 an upper limit on the elastic scattering spin-independent cross
section between the weakly interacting massive particle (WIMP) and
the nucleon~\cite{Ahmed:2009zw}.  Based upon this, they have identified an allowed parameter region of
the WIMP mass of 40~GeV $<m_{\rm DM}<$ 200~GeV which is consistent with both the CDMS II experiment and
the DAMA/LIBRA data.
 We
discuss the implication of this mass constraint  and show that the $^7$Li problem
can still be resolved together with the $^6$Li abundance, but only if the negatively charged particles decay into a lighter
dark-matter particle via a  weak charged boson exchange.

The primordial lithium abundances can be  inferred from measurements of absorption line profiles in metal-poor stars (MPSs).  These stars exhibit  roughly constant values of the
abundance ratio, $^7$Li/H, as a
function of metallicity~\cite{Spite:1982dd,Ryan:1999vr,Melendez:2004ni,Asplund:2005yt,bon2007,shi2007,Aoki:2009ce} implying a primordial abundance of 
$^7$Li/H$=(1-2) \times 10^{-10}$.  The standard BBN
model, however,  predicts a value that is a factor of $2-4$ higher (e.g.,
$^7$Li/H=$(5.24^{+0.71}_{-0.67})\times 10^{-10}$~\cite{Cyburt:2008kw})  when one
uses the baryon-to-photon ratio determined from an analysis~\cite{Dunkley:2008ie} of data from the Wilkinson Microwave Anisotropy Probe (WMAP) of the cosmic
microwave background (CMB) radiation.  This discrepancy
requires a mechanism to reduce  the $^7$Li abundance inferred from BBN.  The combination of 
atomic and turbulent diffusion~\cite{Richard:2004pj,Korn:2007cx} might
have reduced the $^7$Li abundance in stellar atmospheres, but this
possibility has not yet been established~\cite{Lind:2009ta}.  

An even more intriguing result  concerns the $^6$Li/$^7$Li isotopic ratios for MPSs.  These have been determined~\cite{Asplund:2005yt} and a high
$^6$Li abundance of $^6$Li/H$\sim 6\times10^{-12}$ has been 
suggested.  This is $\sim$1000 times higher than the standard BBN
prediction.  One should be cautious, however,  in interpreting these results in that  convective
motion in  stellar atmospheres could cause systematic asymmetries in
the observed stellar line profiles and thereby mimic the presence of
$^6$Li~\cite{Cayrel:2007te}.  Nevertheless,  several  MPSs, continue to exhibit  high $^6$Li
abundances even after carefully correcting for  the convection-triggered line
asymmetries~\cite{Perez:2009ax}.   

Be and B abundances have  also observed in MPSs.  $^9$Be \cite{boe1999,Primas:2000ee,Tan:2008md,Smiljanic:2009dt,Ito:2009uv,Rich:2009gj} and B \cite{dun1997,gar1998,Primas:1998gp,cun2000} abundances appear to increase
roughly linearly as the Fe abundance increases.  The absence of a  plateau in the  abundances of Be and B at low metallicity, however, suggests that these elements are not of primordial origin.

Nonthermal nuclear reactions induced by the decay of exotic particles
have been studied~\cite{decaying,Kusakabe_rad} as a means to  provide a cosmological solution to the Li problems.  Nonthermal reactions triggered by the
radiative decay of long-lived particles can produce $^6$Li nuclides up to
a level $\sim 10$ times larger than the observed level without causing
discrepancies in abundances of other light nuclei or the CMB energy spectrum~\cite{Kusakabe_rad}.

Another solution to the lithium problems of particular interest here is that due to the presence of negatively charged massive particles $X^-$~\cite{cahn:1981,Dimopoulos:1989hk,rujula90} during the BBN epoch.   They affect the nucleosynthesis in a different way~\cite{Pospelov:2006sc,Kohri:2006cn,Cyburt:2006uv,Hamaguchi:2007mp,Bird:2007ge,Kusakabe:2007fu,Kusakabe:2007fv,Jedamzik_nega,Kamimura:2008fx,Pospelov:2007js,Kawasaki:2007xb,Jittoh_nega,Pospelov:2008ta}.  The $X^-$ particles become electromagnetically  bound
to positively charged nuclides with binding energies of $\sim
O(0.1-1)$~MeV with the largest binding energies  for heavier nuclei with larger charges.  Since these binding energies are low, the bound states cannot form until late in the BBN epoch.  At the low temperatures associated with late times, nuclear
reactions  are no longer efficient. Hence, the effect
of the $X^-$ particles is rather small.  Interestingly, however,  the
$X^-$ particles can catalyze the  preferential production of
$^6$Li~\cite{Pospelov:2006sc} along with the  weak destruction of
$^7$Be~\cite{Bird:2007ge,Kusakabe:2007fu}.  

A large enhancement of the
$^6$Li abundance was first suggested~\cite{Pospelov:2006sc} to result from an $X^-$ bound to $^4$He (denoted as $^4$He$_X$).  This 
enables the $X^-$-catalyzed transfer reaction of $^4$He$_X$($d$,$X^-$)$^6$Li,
whose cross section could be seven orders of
magnitude larger than the corresponding BBN $^4$He($d$,$\gamma$)$^6$Li reaction.  The cross section for this reaction, however,  was calculated in a
more rigorous quantum three-body model~\cite{Hamaguchi:2007mp} and shown to be about an order of magnitude
smaller than the estimate adopted in Ref.~\cite{Pospelov:2006sc}.

Additional enhancements in $X^-$-catalyzed transfer reaction rates for the
$^4$He$_X$($t$,$X^-$)$^7$Li, $^4$He$_X$($^3$He,$X^-$)$^7$Be, and
$^6$Li$_X$($p$,$X^-$)$^7$Be reactions were  assumed in Ref.~\cite{Cyburt:2006uv}.  The rates for
those reactions are, however, not as greatly enhanced as that of
the $^4$He$_X$($d$,$X^-$)$^6$Li because they involve  a $\Delta
l=1$ angular momentum transfer  and consequently  a large hindrance of the
nuclear  matrix element~\cite{Kusakabe:2007fu}.  This
 has been confirmed in  recent detailed quantum many-body
calculations~\cite{Kamimura:2008fx}.

The resonant  $^7$Be$_X$($p$,$\gamma$)$^8$B$_X$ reaction
through the first atomic excited state of $^8$B$_X$ was  suggested~\cite{Bird:2007ge} as a means to reduce the primordial $^7$Li
abundance~\footnote{In standard BBN with the baryon-to-photon ratio
inferred from WMAP,  $^7$Li is produced mostly as $^7$Be during the  BBN
epoch.  Subsequently, $^7$Be is transformed into $^7$Li by electron capture.}.  A rate for 
this reaction has been calculated in a rigorous quantum three body model~\cite{Kamimura:2008fx},
which roughly reproduces the value of Ref.~\cite{Bird:2007ge} but is somewhat inefficient 
in destroying $^7$Be$_X$.   The resonant reaction $^7$Be$_X$+$p$ $\rightarrow ^8$B$^*$($1^+$, 0.770 MeV)$_X$ $\rightarrow
^8$B$_X$+$\gamma$ through
the atomic ground state of $^8$B$^\ast$($1^+$,0.770~MeV)$_X$, i.e., an atom
consisting of the $1^+$ nuclear excited state of $^8$B and an $X^-$
particle has also been proposed~\cite{Kusakabe:2007fu} as a process for $^7$Be$_X$ destruction.  From a
more realistic estimate of binding energies between nuclides and $X^-$
particles~\cite{Kusakabe:2007fv}, this
resonant reaction was found to exist, but the resonance energy level is too high to efficiently destroy
$^7$Be$_X$.  

The $^8$Be$_X$+$p$ $\rightarrow ^9$B$_X^{\ast{\rm a}}
\rightarrow ^9$B$_X$+$\gamma$ reaction through the $^9$B$_X^{\ast{\rm a}}$
atomic excited state
of $^9$B$_X$ has also been studied~\cite{Kusakabe:2007fv}.  However this reaction was
found to be not operative because
its resonance energy is relatively large (see Table 2 of Ref.~\cite{Kusakabe:2007fv}).
A resonant reaction
$^8$Be$_X$($n$,$X^-$)$^9$Be through the atomic ground state of
$^9$Be$^\ast$($1/2^+$,1.684~MeV)$_X$, i.e., an atom composed of the $1/2+$
nuclear excited state of $^9$Be and an $X^-$ particle, has also been suggested
as a possible reaction to produce mass 9
nuclides~\cite{Pospelov:2007js}.  Kamimura et al.~\cite{Kamimura:2008fx}, however,  adopted a root mean square charge
radius for  $^8$Be of 3.39~fm as a more realistic input.  They then found that $^9$Be$^*$($1/2^+$, 1.684~MeV)$_X$ is not a
resonance but a bound state located below the $^8$Be$_X$+$n$
threshold.  The resonant $^8$Be$_X$($n$,$X^-$)$^9$Be$_X$ reaction is
thus not likely to contribute.

Neutral $X$-nuclei, i.e., $p_X$,
$d_X$ and $t_X$ have also been suggested~\cite{Jedamzik_nega}  as a means to produce and
destroy Li and Be through two $\alpha$-induced $X^-$ stripping  reactions
$d_X$($\alpha$,$X^-$)$^6$Li and $t_X$($\alpha$,$X^-$)$^7$Li, and three
$p_X$ induced stripping reactions $p_X$($^6$Li,$^3$He$\alpha$)$X^-$,
$p_X$($^7$Li,2$\alpha$)$X^-$ and
$p_X$($^7$Be,$^8$B)$X^-$.  The result, however,  relies on
reaction rates calculated within the framework of the Born
approximation, which is a poor approximation in this low-energy regime~\cite{Dimopoulos:1989hk,Kamimura:2008fx}.  The rates
for those reactions and those for charge-exchange reactions of
$p_X$($\alpha$,$p$)$\alpha_X$, $d_X$($\alpha$,$d$)$\alpha_X$ and
$t_X$($\alpha$,$t$)$\alpha_X$ have been calculated in a rigorous dynamical  quantum many-body
treatment in Ref.~\cite{Kamimura:2008fx}.  They found
that the cross sections for  the charge-exchange reactions were much larger than
those of the nuclear reactions, and that the neutral bound states $p_X$, $d_X$
and $t_X$ were  immediately changed to $\alpha_X$ before they could react with ambient
nuclei.  The late time production and destruction of Li and Be, therefore,
do not significantly affect the BBN  as shown in this
letter.  

The solution to the
lithium problems in this catalyzed BBN model have  been explored  by solving the full
Boltzmann equations for the recombination and the ionization of nuclides and
$X^-$ particles coupled to the nuclear
reactions~\cite{Kusakabe:2007fu,Kusakabe:2007fv,Jedamzik_nega}.
Constraints
on specific supersymmetric
models through the catalyzed BBN calculation have been 
obtained in Refs.~\cite{Cyburt:2006uv,Kawasaki:2007xb,Jittoh_nega,Kawasaki:2008qe,Bailly:2008yy}.

%\section{Model}The leptonic $X^-$ particle is assumed to exist in the early universe.
Candidates for the  leptonic $X^-$ particle of interest in these models are the spin 0 supersymmetric partners of the standard-model leptons.  Such $X^-$ particles (and their
antiparticles $X^+$) would be produced copiously  in the hot early universe and subsequently annihilate.  Their
annihilations, however, would  freezeout at some epoch.  The residual $X^+$ particles do not
affect BBN because they do not bind to the positively-charged nuclei.  
It is possible, however,  that the decay of both $X^\pm$ particles affect the final
 light element abundances through electromagnetic and/or
hadronic showers.  Here, however, we only consider
the $X$-nuclear reactions, and not the effect of subsequent decay.

%\subsection{Nuclear Binding Energies}
The binding energies of nuclei bound
to $X^-$ particles, i.e., $X$-nuclei, have been derived by
taking account of the modified Coulomb interaction with the
nucleus~\cite{Kusakabe:2007fv} under the assumption that the  mass of the $X^-$ particle is 
much heavier than the nucleon mass.  

%\subsection{Reaction Network}
We performed the detailed network calculation of the catalyzed BBN~\cite{Kusakabe:2007fv} taking
into account  both the recombination and ionization of $X^-$ particles with
nuclei and thermonuclear reactions and
$\beta$-decay of normal nuclides and $X$-nuclei.   We adopt all of the new reaction
rates from the rigorous  quantum many-body dynamical calculations of Ref.~\cite{Kamimura:2008fx}.  For the
$^7$Be$_X$($p$,$\gamma$)$^8$B$_X$ resonant reaction through an atomic
excited state of
$^8$B$_X$, rates for different masses of $X^-$ have been
published~\cite{Kamimura:2008fx}.  For our purposes we
adopt their rate for an infinite $X^-$ mass.  Our
results are thus completely different from  previous studies without the use of the new cross
sections.

%\subsection{Constraints on the Primordial $^6$Li Abundance}
We adopt the  constraint on the primordial $^6$Li abundance from the observations
in MPSs.  The primordial abundances of $^6$Li and $^7$Li could be higher than the
observed abundances~\cite{Asplund:2005yt} considering the possible
effect of stellar depletion of initial surface abundances.  The observed
abundances should, therefore, be considered a lower limit to the true primordial ones.  
Since $^6$Li is more fragile to nuclear burning than $^7$Li
~\cite{Richard:2004pj}, its depletion factors could be
 larger than those for $^7$Li.  We adopt a conservative limit of a factor of 10
above the mean value of ($^6$Li/H)$_{\rm MPS}=(7.1\pm0.7)\times
10^{-12}$~\cite{Asplund:2005yt}, and a 3~$\sigma$ lower limit to the mean value
times a factor of $1/3$.  The limit on  the $^6$Li/H abundance are  thus $1.7\times
10^{-12}\le ^6$Li/H $\le 7.1\times 10^{-11}$.

\begin{figure}[tbp]
\begin{center}
\includegraphics[width=8.0cm,clip]{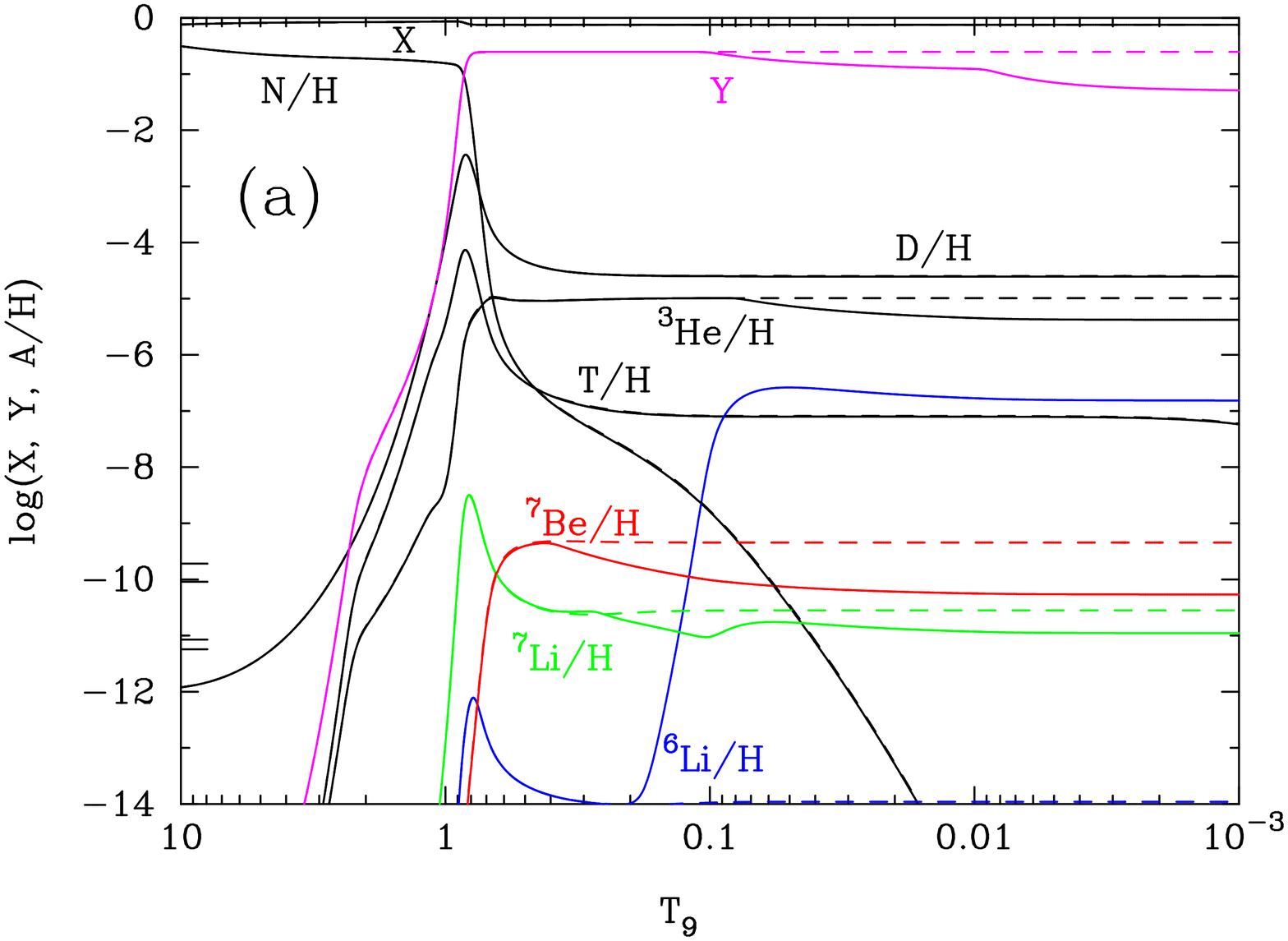}
\includegraphics[width=8.0cm,clip]{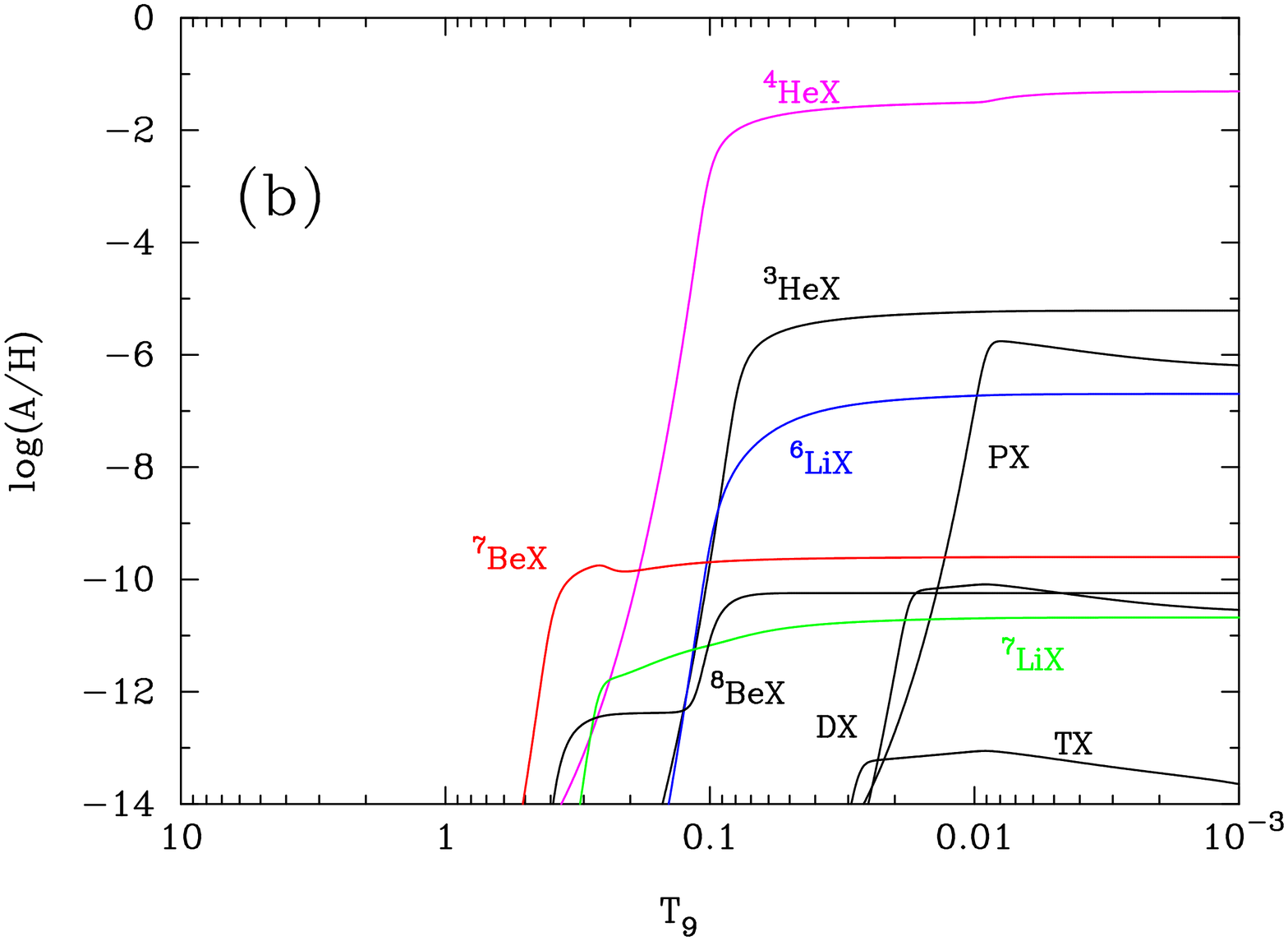}
\end{center}
\caption{(color online). Calculated abundances of normal nuclei (a) and $X$-nuclei (b)
 as a function of $T_9$ (solid lines).  The abundance and the lifetime
 of the $X^-$ particle are set to be $Y_X=n_X/n_b=0.05$ and $\tau_X=\infty$,
 respectively. The dashed lines correspond to the standard BBN case.~\label{fig1}}
\end{figure}

%\section{Results}
%\subsection{BBN Calculation Result}
Figures \ref{fig1}a and \ref{fig1}b show the results of a catalyzed BBN
calculation.  For these figures the $X^-$ abundance was taken to be 5\% of the total baryon
abundance,  i.e. $Y_X=n_X/n_b=0.05$, where $n_X$ and $n_b$ are the
number densities of the $X^-$ particles and baryons, respectively.
Figure~\ref{fig1}a shows the evolution of  normal nuclei while Figure~\ref{fig1}b
corresponds to $X$-nuclei.

The abundances of the normal nuclei are very similar to the standard BBN
abundances until the temperature reaches $T_9\sim 0.5$.  The $X^-$
particles then combine with $^7$Be at $T_9 \sim 0.5$ and subsequently $^7$Li at $T_9
\sim 0.3$.  The $^7$Be$_X$ produced by these $X^-$  captures (Fig.\
\ref{fig1}b) is then destroyed by the $^7$Be$_X$($p$,$\gamma$)$^8$B$_X$ reaction, primarily through
the atomic excited state of $^8$B$_X$~\cite{Bird:2007ge}, and
secondarily  through the atomic ground state $^8$B$^*$($1^+$, 0.770
MeV)$_X$ composed of the $^8$B$^*$($1^+$, 0.770
MeV) nuclear excited state
and an $X^-$ particle~\cite{Kusakabe:2007fu}.  We have assumed that $^8$B$_X$ inter-converts to
$^8$Be$^*(2^+$,3~MeV)$_X$ by $\beta$-decay with a rate given by the normal  $^8$B
$\beta$-decay rate multiplied by a correction term $(Q_X/Q)^5$, where $Q$
and $Q_X$ are
the $Q$-values of the standard $\beta$-decay and that of $\beta$-decay for
$X$-nuclei~\cite{Kusakabe:2007fv}.  The produced $^8$Be$^*(2^+$,3~MeV)$_X$ then immediately decays to the
three-body channel $\alpha$+$\alpha$+$X^-$~\cite{Kamimura:2008fx}.

When
the temperature decreases to $T_9 \sim
0.1$, the $X^-$ particles bind to $^4$He.  Then, the $X^-$-catalyzed
transfer reaction $^4$He$_X$($d$,$X^-$) operates to produce normal
$^6$Li and $^6$Li$_X$ (after the recombination).   Because
of the small binding energies to the  $X^-$ (see Table I of Ref.~\cite{Kusakabe:2007fv}), neutral $X$-nuclei do not form until  late times corresponding to  $T_9\sim0.03$
(for $t_X$), $T_9\sim0.02$ (for $d_X$) and $T_9\sim0.01$ (for $p_X$).  The neutral
$X$-nuclei then mainly react with $^4$He nuclei to lose their $X^-$ and to produce
$^4$He$_X$ (as a result of the precise calculation~\cite{Kamimura:2008fx}) so that abundances of neutral $X$-nuclei are kept low.  Nuclear
reactions triggered by neutral $X$-nuclei are thus not important.

\begin{figure}[tb]
\begin{center}
\includegraphics[width=8.0cm,clip]{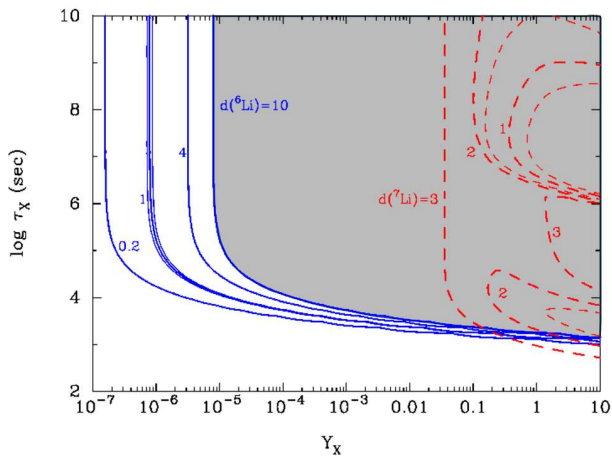}
\end{center}
\caption{(color online). Contours of constant lithium isotopic abundances relative to
 observed values in MPSs, i.e., $d$($^6$Li) = $^6$Li$^{\rm
 Calc}$/$^6$Li$^{\rm Obs}$ (solid curves) and $d$($^7$Li) = $^7$Li$^{\rm
 Calc}$/$^7$Li$^{\rm Obs}$ (dashed curves).  The adopted observational abundances are
 $^7$Li/H$= (1.23^{+0.68}_{-0.32})\times 10^{-10}$~\cite{Ryan:1999vr}
 and $^6$Li/H$=(7.1\pm 0.7)\times 10^{-12}$~\cite{Asplund:2005yt}.  Thin
 solid and dashed lines around the lines of $d$($^{6,7}$Li) = 1
 correspond to the 1~$\sigma$ uncertainties in the observational
 constraint.  The gray region is observationally excluded by the overproduction of
 $^6$Li.~\label{fig2}}
\end{figure}

%\subsection{Observational Constraints on the $X^-$ Abundance}
Figure\ \ref{fig2} shows the contours of $d$($^6$Li) (solid curves) and
$d$($^7$Li) (dashed curves) in the parameter plane of the abundance
$Y_X$ and the lifetime $\tau_X$ of the $X^-$ particles.  $d$($^A$Li)=$^A$Li$^{\rm
 Calc}$/$^A$Li$^{\rm Obs}$ is the ratio of the calculated abundance to the
 observed abundance.  The solid curves
labeled $d$($^6$Li)=10 and 0.2 correspond to  upper and lower limits on
the abundance constraint ($1.7\times 10^{-12} \le ^6$Li/H $\le 7.1\times
10^{-11}$).  Above the $d$($^6$Li)=0.2 line,  $^6$Li is produced
at the observed level in MPSs.  

The thick dashed curves are for $d$($^7$Li)=1, 2 and 3.  
The dashed curve for $d$($^7$Li)=2 intersects the contours of $d$($^6$Li) for
$Y_X\gtrsim 1$ and $\tau_X\approx (1-2)\times 10^{-3}$.  In the region above this curve the $^7$Li abundance is lower than $^7$Li/H$\approx
2.5 \times 10^{-10}$.  The updated parameter region for a simultaneous  solution to
the $^6$Li and $^7$Li abundances in BBN with negatively charged particle is:
$Y_X \gtrsim 1$ and $\tau_X\approx (1-2)\times 10^3$~s.

For $\tau_X \gtrsim 10^4$~s and $Y_X \gtrsim 0.3$, the calculated abundance of
$^7$Li increases slightly due to  the $^4$He$_X$($t$,$X^-$)$^7$Li and
$^4$He$_X$($^3$He,$X^-$)$^7$Be reactions which produce some amount of $^7$Li.
However, this parameter region is not allowed  due to an extreme overproduction of  $^6$Li.  The gray region in Fig.\ \ref{fig2} is the parameter region excluded by the overproduction of $^6$Li.

If the $X^-$ particle decays via the  weak interaction, $^7$Be$_X$ converts to $^7$Li  by a weak
charged current transition from $X^-$ to $X^0$, i.e.  $^7$Be$_X
\rightarrow ^7$Li+$X^0$~\cite{Jittoh_nega,Bird:2007ge}.  This case is shown in Fig. 6 of Ref.~\cite{Kusakabe:2007fv}.  The results from that study  do not change by
the implementation of the new cross sections adopted here.
The larger destruction rate associated with the $^7$Be$_X \rightarrow
^7$Li+$X^0$ decay followed by the $^7$Li($p$,$\alpha$)$^4$He and
$^7$Li($X^-$,$\gamma$)$^7$Li$_X$($p$,$2\alpha$)$X^-$
reactions~\cite{Kusakabe:2007fv,Bird:2007ge} or a further conversion of $^7$Li$_X$ by the
weak interaction~\cite{Jittoh_nega} shifts the contours of the $^7$Li abundance toward smaller values of $Y_X$.  In this case the parameter region which solves both the
$^6$Li and $^7$Li problems is $Y_X \approx 0.04-0.2$ and $\tau_X \approx (1.4-2.6) \times 10^3$~s.

%\section{Discussion}
We now consider a model in which the present cold dark matter (DM) was produced by the
decay of $X^\pm$ particles, i.e., $Y_{\rm DM} \geq Y_X$.  The WMAP-CMB constraint on the cosmological
density parameter of cold DM $\Omega_{\rm CDM}=0.2$
then corresponds to $m_{\rm DM} Y_{\rm DM} \leq 4.5$~GeV.  The constraints on
$Y_X$ required to resolve the $^6$Li and $^7$Li problems then imply a range for the the dark-matter mass $m_{\rm DM}$.  In the case where the reactions $^7$Be$_X$+$p \rightarrow ^8$B$_X^{\ast {\rm a}}
\rightarrow ^8$B$_X$+$\gamma$~\cite{Bird:2007ge} and $^7$Be$_X+p\rightarrow
^8$B$^*$($1^+$, 0.770 MeV)$_X \rightarrow ^8$B$_X+\gamma$~\cite{Kusakabe:2007fu}
destroy $^7$Be$_X$, the DM mass is thus constrained to be $m_{\rm DM}
\leq 4.5$~GeV.  On the other hand, when the $^7$Be$_X \rightarrow
^7$Li+$X^0$ reaction~\cite{Jittoh_nega,Bird:2007ge} is included, the allowed
mass range increases  to $m_{\rm DM} \leq 20-110$GeV.

Comparing this result to the allowed parameter region for the DM mass of
40~GeV $<m_{\rm DM}<$ 200~GeV~\cite{Ahmed:2009zw}, implies  that only an  $X^-$
particle which decay via the  weak interaction can have existed
with sufficient abundance  to reduce $^7$Li produced from BBN.  On the other
hand, if the $X^-$ particles  do not  decay via the  weak interaction they are excluded.

%\section{Conclusions}
In summary, we have  re-investigated BBN in the presence of negatively-charged massive
particles $X^-$ by solving the rate equations with an improved  nuclear reaction network
code~\cite{Kusakabe:2007fv}.  We have adopted  the newest quantum many-body dynamical
calculations~\cite{Kamimura:2008fx}.  With the new rates, we find that, contrary to the speculation in  previous studies, the neutral $X$-nuclei, i.e., $p_X$, $d_X$ and $t_X$,
do not significantly affect the BBN abundances. Furthermore, based upon
constraints for the mass of DM particles from the possible CDMS-II
events we conclude that only   $X^-$ particles which decay
into the DM via the  weak
interaction can simultaneously reduce $^7$Li to the desired level while producing enough $^6$Li in the early universe.  

%\section{Note}
Finally, we note  that in this revised catalyzed BBN model 
there is no signature in the abundances of nuclei heavier than Be.
Thus, if one were to find a primordial plateau abundance of Be or B, it
would require an origin other than this catalyzed BBN model.  There are three
physical processes by which enhanced  abundances of nuclei heavier
than Be could have been formed.  The first is the
cosmological cosmic ray nucleosynthesis induced by supernova explosions
during  an early epoch of structure formation~\cite{Rollinde:2004kz}.  This
model can resolve the $^6$Li problem and leave possible abundance plateaus
of $^9$Be and B~\cite{kus07a,rol2008}.  The second is the BBN model
including a long-lived strongly interacting massive
particle.  Signatures of such particles are possibly
left on the primordial abundances of Be and B which may be found in
future astronomical observations of MPSs~\cite{Kusakabe:2009jt}.  The third is
the inhomogeneous BBN model which can lead to a  high primordial abundance of
$^9$Be~\cite{IBBN}.

\begin{acknowledgments}
This work was supported by Grants-in-Aid for JSPS Fellows (21.6817), for
 Scientific Research of JSPS (20244035), and for Scientific Research on
 Innovative Area of MEXT (20105004), the Mitsubishi Foundation, and JSPS
 Core-to-Core Program EFES. Work at the University of Notre Dame was
 supported by the U.S. Department of Energy under Nuclear Theory Grant
 DE-FG02-95-ER40934.
\end{acknowledgments}

%\bibliography{ref_nega3}

%%%%%%%%%%%%%%%%%%%%%%%%%%%%%%%%%%%%%%%%%%%%%%%%%%%%%%

\end{document}